\documentstyle{lamuphys}
\input{psfig}

\def\deg   {$^\circ$}
\def\phiI  {$\varphi_i$}

\def\Pcos  {\ifmmode{\varphi^u}   \else{$\varphi^u$}\fi}
\def\Jcos  {\ifmmode{J_{-21}^u}\else{$J_{-21}^u$}\fi}
\def\NH    {\rm N_{\scriptscriptstyle H}}
\def\Ha    {${\rm H}\alpha$}

\def\eg    {{\it e.g.,\ }}

\def\qv    {{\it q.v.,\ }}
\def\etal  {{\it\ et al.}}
\def\intensity{\ifmmode{{\rm\ erg\ cm}^{-2}{\rm\ s}^{-1}
      {\rm\ Hz}^{-1}{\rm\ sr}^{-1}}
      \else {\ erg cm$^{-2}$ s$^{-1}$ Hz$^{-1}$ sr$^{-1}$}\fi}

\def\flux{\ifmmode{{\rm erg\ cm}^{-2}{\rm\ s}^{-1}}\else {erg
cm$^{-2}$ s$^{-1}$}\fi}

\def\fluxdensity{\ifmmode{{\rm erg\ cm^{-2}\ s^{-1}\ Hz^{-1}}}\else {erg
cm$^{-2}$ s$^{-1}$ Hz$^{-1}$}\fi}
\def\phoflux{\ifmmode{{\rm\ phot\ cm}^{-2}{\rm\ s}^{-1}}\else {\ phot
cm$^{-2}$ s$^{-1}$}\fi}
\def\eflux{\ifmmode{{\rm\ erg\ cm}^{-2}{\rm\ s}^{-1}}\else {\ erg
cm$^{-2}$ s$^{-1}$}\fi}

\def\nH {\ifmmode{\rm n_{\scriptscriptstyle H}}\else{n$_{\scriptscriptstyle H}$}\fi}
\def\NH {\ifmmode{{\rm N_{\scriptscriptstyle H}}}\else{N$_{\scriptscriptstyle H}$}\fi}
\def\Np {\ifmmode{{\rm N_{\scriptscriptstyle p}}}\else{N$_{\scriptscriptstyle p}$}\fi}

\def\ne {\ifmmode{\rm n_{\scriptscriptstyle e}}\else{n$_{\scriptscriptstyle e}$}\fi}
\def\np {\ifmmode{\rm n_{\scriptscriptstyle p}}\else{n$_{\scriptscriptstyle p}$}\fi}

\makeatletter
\let\chapter\hid@chapter
\makeatother
\newcommand{\T}{ {\scriptscriptstyle {\rm T}} }
\newcommand{\Tdec}{{\hbox
      {$\displaystyle\left({\delta T \over T}\right)_r$} }}

\begin{document}
\pagenumbering{arabic}

\title{The Metagalactic Ionizing Field in the Local Group}

\author{~~~J. Bland-Hawthorn}

\institute{Anglo-Australian Observatory, P.O. Box 296,
Epping, NSW 2121, Australia.}

\maketitle

\begin{abstract}
We discuss the sources which are likely to dominate the
ionizing field throughout the Local Group. In terms 
of the limiting flux to produce detectable \Ha\ 
emission ($\sim 4-10\times 10^3$\phoflux),
the four dominant galaxies (M31, Galaxy, M33, LMC) have
spheres of influence which occupy a small fraction ($5-10$\%)
of the Local Volume. There are at least two possible
sources of ionization whose influence could be far more
pervasive: (i) a cosmic background of ionizing photons;
(ii) a pervasive warm plasma throughout the Local Group. 
The {\sl COBE} FIRAS sky temperature measurements permit a 
wide variety of plasmas with detectable ionizing fields.  
It has been suggested (Blitz\etal\ 1996; Spergel\etal\ 1996;
Sembach\etal\ 1995, 1998) that a substantial fraction of
high velocity clouds are external to the Galaxy but within
the Local Group. Deep \Ha\ detections are the crucial
test of these claims and, indeed, provide a test bed for
the putative Local Group corona.  
\footnote{Invited talk, to appear in the Proceedings of the ESO/ATNF Workshop
{\it Looking Deep in the Southern Sky}, 10-12 December 1997, Sydney, Australia,
Eds. R. Morganti  and W. Couch}
\end{abstract}

\section{Introduction}

In keeping with this Workshop, we concentrate on what one can 
hope to learn, in the coming decade, from deep spectroscopic 
studies of diffuse line
emission.  Such techniques are now used by several groups in
both hemispheres.  (For progress to date, see Reynolds\etal\ 1997;
Bland-Hawthorn 1997.) The spectroscopic methods (e.g. `staring') 
achieve extremely deep levels ($\S$2) and can detect the presence
of extremely weak ionizing fields.

To encourage a broader campaign of spectroscopic studies,
we deduce the expected level of ionizing flux within the 
Local Group. We provide an inventory of possible sources, 
and discuss the prospect of a ubiquitous warm plasma.  
The discovery of such a medium would have fundamental
implications:

\smallskip
\hangindent \parindent \hangafter 1
$\bullet$
A recent review of the cosmic baryon budget finds that 
a large fraction of `missing' baryons may well be tied up in
warm gas within galaxy groups (Fukugita,
Hogan \& Peebles 1998). To be consistent with weak x-ray 
detections (Pildis, Bregman \& Evrard 1995), the diffuse 
gas needs to radiate predominantly at EUV wavelengths.

\hangindent \parindent \hangafter 1
$\bullet$
Such an ionizing field could render an
HI cloud optically visible everywhere within the
Local Group. A truly {\it cosmic} ionizing field
could render the same cloud visible over cosmological
distances (Circovic\etal\ 1998). 

\hangindent \parindent \hangafter 1
$\bullet$
The existence of warm, tenuous gas in loose galaxy
groups has important ramifications for the Ly$\alpha$
forest at low redshift detected along QSO sight lines
(Morris\etal\ 1991; Bahcall\etal\ 1991).

\hangindent \parindent \hangafter 1
$\bullet$
An intragroup medium would indicate the presence
of Galactic fountains or winds, or even primordial 
gas dating back to the formation of the group. If
confirmed, this medium is expected to radically influence 
the star-formation history of the Local Group 
(e.g. Dressler 1986; van den Bergh 1994).

\section{The deepest spectroscopic detections}

The deepest spectroscopic limits are achieved by Fabry-Perot 
`staring'. Several groups have shown what is possible with
modern day optics and detectors (\qv Bland-Hawthorn 1997).
The power of the method comes from dispersing the light of
a single emission line onto $\sim5-10$\% of a wide area CCD. 
A hard experimental limit is about 1 mR (1$\sigma$) at 1\AA\
resolution which is 8 mag below sky within that narrow
band. This is close to the experimental limit of space-borne
observations in broad optical bands (Vogeley 1998).
Both broad and narrow band limits fall far below the 
zodiacal light level arising from interplanetary dust and gas.

For our deep limit, what is the corresponding ionizing flux?
From the surface of an HI slab optically
thick to ionizing photons, the emission
measure is ${\cal E}_m = \int n_e n_{H^+}\;dl =n_e n_{H^+} L\
{\rm cm^{-6}\; pc}$ where $L$ is the thickness of the ionised region.
The resulting emission measure for an ionizing flux \phiI\ is then
${\cal E}_m = 1.25\times 10^{-2} \varphi_4 \ {\rm cm^{-6}\; pc}$
($\equiv 4.5 \varphi_4 \ {\rm mR}$) where
$\varphi_i = 10^4 \varphi_4$. The 1 mR spectroscopic limit corresponds 
to an ionizing flux of less than 10$^4$\phoflux. 
To put this into perspective, this is the expected 
level of \Ha\ recombination emission from the interstellar
medium {\it within the Solar System} due to the mean Galactic 
ionizing field (Reynolds 1984)!

\section{Cosmic UV background}

From \Ha\ non-detections towards extragalactic HI,
the current best 2$\sigma$ upper limit (Vogel\etal\ 1995) for the
universal ionizing background is \Jcos\ $<$ 0.08 (\Pcos\ 
$< 2-4\times 10^4$\phoflux).  \Jcos\ is the ionizing flux density of the
cosmic background at the Lyman limit in units of 10$^{-21}$ erg
cm$^{-2}$ s$^{-1}$ Hz$^{-1}$ sr$^{-1}$; \Pcos\ ($=\pi$\Jcos$/h$) is the
equivalent photon flux at face of a uniform, optically thick slab.
The unexpectedly high number of Ly$\alpha$ absorbers towards 3C273
gives \Jcos\ $\approx$ 0.006 from the proximity effect 
(Kulkarni \& Fall 1995), although this value is uncertain by at least 
a factor of 5. Sciama (1993; 1998) has suggested a somewhat stronger
ionizing field permeates the Universe due to decaying tau neutrinos 
with masses 24 eV and lifetimes $\sim$10$^{23}$ s.

\begin{figure}
\label{fig1}
\protect\centerline{
\psfig{figure=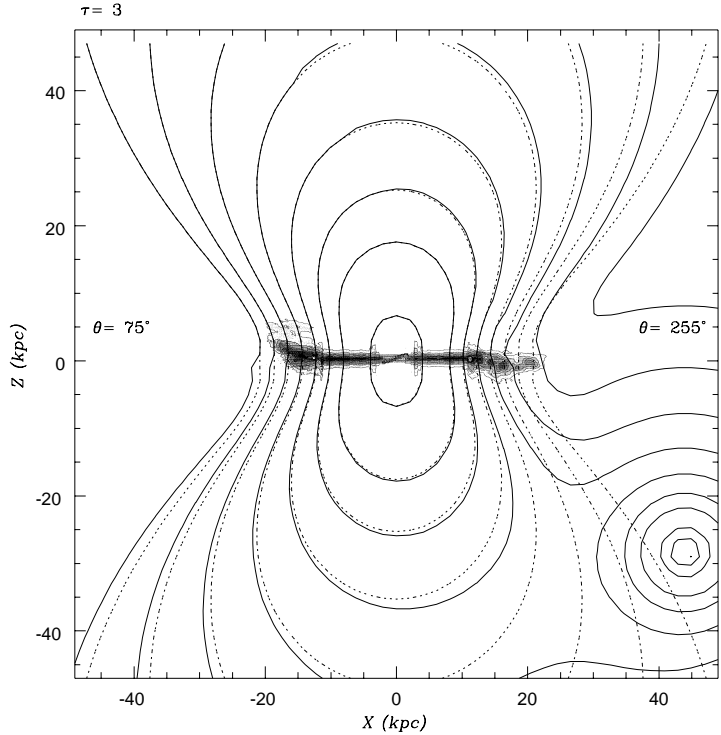,height=4truein,width=4truein}}
\caption[] {The Galactic halo ionizing field. The coordinates are
with respect to a plane perpendicular to the Galactic disk (with
the Galactic Centre at the origin) at a constant galactic azimuth
angle (75\deg, 255\deg). The dotted lines show the ionizing flux
($\varphi_4$) due to the galactic disk; the solid lines include the
contribution from the LMC. The opacity of the HI disk (shown in
half tone) has been included. The contours, from outside in, are
for $\log \varphi_4 = 1, 1.25, 1.5, 1.75, 2, 2.25, 2.5, 3$\phoflux.
The minor contribution from the Galactic corona is omitted (BM).}
\end{figure}

\section{An inventory of the Local Group}

The Local Group is a member of the Coma-Sculptor Cloud within the Local 
Supercluster, a highly flattened structure extending over more than
10 Mpc (Tully 1988). Outside the Local Group, the closest structures
are the Sculptor, Maffei/IC342, and M81 groups. While the Maffei group
has the biggest dynamical influence on us, the Sculptor group is
expected to produce the stronger ionizing field at the periphery of
our group.

The Local Group has at least 40 members (see appendix; Tully 1988). 
More than 80\% of the visual light emerges from the great rajahs, M31 
and the Galaxy. This increases to 95\% if we include M33 and the LMC.
The remaining subjects are dwarf irregulars, spheroidals or ellipticals
which congregate around the royal members, or form small associations 
near the edge of the group. For convenience, we define the Local Volume 
as a sphere extending to 1 Mpc radius from the Local Group barycentre (LGB).
\footnote{R.B. Tully (1998, private communication) points out that a 
criterion for group membership requires total or continuing collapse 
towards the common potential of the group. However, the distances,
space motions, and the underlying mass distribution are not well enough 
known to be sure of which of the outermost objects are bound.}

\subsection{Young stars}

Our model of the ionizing background in the Local Group includes 
the poloidal radiation fields of the Galaxy, M31 and M33. The Galaxy halo 
includes the contribution from the LMC (see Fig.~\ref{fig1}).  To a good 
approximation, the UV radiation field arising from an opaque, stellar disk 
is given by (Bland-Hawthorn \& Maloney 1998; hereafter BM)
\begin{equation}
\label{uvdisk}
\varphi^\star = \varphi^\star_o \ e^{-\tau}\ 
{\rm r}_{100}^{-2} \cos^{0.6\tau+0.5}
\theta \ \ \ \ \ \ \ \ \phoflux .
\end{equation}
The polar angle $\theta$ ($0 \leq \theta < \frac{\pi}{2}$) is
measured from the spin axis, $r_{100}$ is the radius in units
of 100 kpc, and the Lyman limit optical depth is $\tau \leq 10$. 
In order to explain Magellanic Stream \Ha\ detections and OB star counts, 
our model favours $\varphi^\star_o = 2.8\times 10^6\phoflux$ and 
$\tau \approx 2.8$ for the Galaxy.
For the purposes of this Workshop, the ionizing fields for M31 and M33 
are arbitrarily scaled with respect to the Galaxy by their blue 
luminosities for the same opacity. This crude assumption may well be 
flawed since the evolutionary histories of these galaxies could be quite 
different.

While the ionizing fields are expected to be weak, we include the 
contribution of the dwarfs for completeness. Deep HI surveys may have
found gas linked to dwarf galaxies (\qv Young \& Lo 1997)
and even a very weak field could render such gas optically visible. 
Deeper surveys already in progress are expected to turn up more HI 
associations (Staveley-Smith\etal\ 1996; Briggs\etal\ 1997).

For the dwarf irregulars,
on-going star formation appears confined to the outermost systems,
DDO 210 and Phoenix, which may be evidence enough for a diffuse medium
throughout the Local Group. The lack of star formation in the more central
dwarfs is a possible manifestation of ram-pressure stripping
(van den Bergh 1994; Hirashita, Kamaya \& Mineshige 1997). 

In dwarf spheroidals, there is evidence for a `UV upturn'
presumably from horizontal branch (HB) stars. 
In ellipticals (and S0s), the `UV upturn' population are 
thought comprise hot HB, post-HB star and
post-AGB stars (Dorman 1997; Brown\etal\ 1997). Our model
includes a weak isotropic source for each of the dwarf irregulars
and spheroidals.

\subsection{Galactic coronae}

The depth of x-ray shadows observed towards high-latitude gas clouds 
is consistent with a 0.25 keV patchy coronal halo encompassing the Galaxy
(Burrows \& Mendenhall 1991; Snowden\etal\ 1991; Snowden, McCammon \& 
Verter 1993). Soft x-ray haloes are also observed in a significant
fraction of spirals (Bregman \& Pildis 1994;
Vogler, Pietsch \& Kahabka 1996; cf. Kim\etal\ 1996).

To explore the effect of a hot galactic corona, we assume that the gas 
distribution is described by a non-singular isothermal sphere, with a
density law
\begin{equation}
\label{dens}
n(r) = {n_c\over{(1+r^2/r_c^2)}}\ {\rm cm^{-3}}
\end{equation}
where $r_c$ is the core radius and $n_c$ is the core gas density. For
hot gas temperatures $T_e\geq 2$ keV, the emission is dominated by
thermal bremsstrahlung.  Only photons in the energy
range 13.6 eV to $\sim$250 eV are important to the ionization state of
the gas for the column densities of relevance here.  The ionizing photon 
flux on the inner face of a cloud at radius $r$ (for $r/r_c\le 12$) in
kpc is given by (BM)
\begin{equation}
\label{bremss}
\varphi^{\rm gc}(r) \approx 18 n^2_{-3} r_c {\cal C}(T_e) 
T^{-0.225}_{\rm keV}
\left[{{\eta+1.3(r/r_c)^{1.35}}\over{(1+r^2/r_c^2)^{1.5}}}\right]
\ \ \ \ \phoflux
\end{equation}
where 10$^{-3}n_{-3}$ cm$^{-3}$ is the core gas density. For an
optically thin Galactic disk, $\eta = \pi/4$; when the disk is
completely opaque to ionizing photons, $\eta = 0$, which reduces the
ionizing photon flux by less than a factor of two. 

For the Galaxy, absorption lines towards stars in the Magellanic
Clouds have been used to infer coronal temperatures and gas densities;
for example, Songaila (1981) estimates $T_e\approx 0.05-0.3$ keV
and $n_c\sim 3-7 \times 10^{-4}$ cm$^{-3}$.  
We adopt $r/r_c = 2$, n$_{-3} = 2$, and assume 
$T_e\approx 0.2$ keV, the virial temperature 
of the Galactic halo (for an assumed circular velocity of 220 km s$^{-1}$). 
This density and temperature distribution matches the emission measure and
x-ray luminosity of the diffuse component determined by Wang \& McCray
(1993).

At such low temperatures ($T_{\rm keV}$ $<$ 2), eq.~\ref{bremss} does 
not include the strong EUV ionizing lines produced by the gas, and 
therefore considerably underestimates the true photon flux. We estimate 
the increase in the effective emissivity using Table 4 in Gaetz \& Salpeter 
(1983). Representative correction factors ${\cal C}$ are 28 ($T_e = 0.34$ keV) 
and 55 ($T_e = 0.22$ keV). The cooling time for the gas is $2-3\times 10^8$ 
yrs.  At the lower temperature, the expected ionizing 
flux at 20 kpc radius is $\varphi^{\rm gc} \sim 2\times 10^4$\phoflux.

\subsection{Local Group corona}

The extremely accurate blackbody form of the cosmic microwave
background (CMB) sets an important constraint on an ionised
intergalactic medium. The presence of a hot plasma distorts
the microwave background through Compton scattering of the CMB
photons (Sunyaev \& Zel'dovich 1969).  

The departures from blackbody are quantified by the $y$-parameter such that
\begin{equation}
\label{compy}
y = \int_0^t {{k(T_e-T_r)}\over{m_e c^2}}\sigma_T n_e\ c\ dt .
\end{equation}
where $\sigma_T$ is the Thomson scattering cross section and
$m_e$ is the electron mass. The integration is performed over
the time taken for the photon to traverse the ionised medium.
In most cases, $T_e$ is much greater than the radiation temperature
$T_r$. The expected temperature decrement is then
\begin{equation}
\label{Tdec}
\Tdec = -2 {k T_e \over m_e c^2}\sigma_\T N_e(\mu),
\end{equation}
where $N_e$ is the electron column at an angle $\Theta_c$
($\mu = \cos\Theta_c$) to the LGB sight line.
The multipoles, after expanding the sky temperature in spherical
harmonics as a function of angular position, are easily derived
(Suto\etal\ 1996). The monopole term reduces to
\begin{equation}
\label{T0}
T_0 = \pi \Theta_c
     \,\sigma_\T {k T_e \over m_e c^2} {n_o R_c^2 \over x_o}
\end{equation}
where $n_o$ is the central density of the Local Group corona, 
$x_o$ is the distance of the Galaxy from the centre, $R_c$ is 
the core radius of the corona, and $\Theta_c \equiv \tan^{-1}(x_o/R_c)$.

The {\sl COBE} FIRAS data (Mather\etal\ 1994) imply that the 
Compton $y$-parameter is less than $2.5\times10^{-5}$ (95\%
confidence level).  The upper limit translates to (Suto\etal\ 1996)
\begin{equation}
\label{suto}
n_o R_c^2/x_o\ \ <\ \ 1.1\times10^{22} T^{-1}_{\rm keV}
 \left({1.17 \over \Theta_c}\right)
 \left({y \over 2.5\times10^{-5}}\right)
            \, {\rm cm}^{-2}.
\end{equation}
We adopt an identical form to eq.~\ref{bremss} for the Local Group flux,
and associate $r$ with $x_o$ such that
\begin{equation}
\label{Compy}
\varphi^{lg}(\Theta_c)\ \  <\ \   5\times 10^4 {\cal C}(T_e) 
T^{-1.2}_{\rm keV} 
n_{-3}(\Theta_c) 
{\cal F}(\Theta_c) \left({y \over 2.5\times10^{-5}}\right)
\ \ \ \phoflux
\end{equation}
for which
\begin{equation}
\label{angdep}
{\cal F}(\Theta_c) =  1.17 
 \cos^2\Theta_c
\left(0.8+1.3\tan^{1.35}\Theta_c\right)
\left({\sin\Theta_c \over \Theta_c}\right)
\end{equation}
The angular term ${\cal F}(\Theta_c)$ is well behaved,
with a value of 0.94 towards
LGB, peaking at 1.2 near $\Theta_c = 0.55$, falling
to 0.75 at Suto's canonical value of $\Theta_c = 1.17$, and
to zero beyond here.

After Suto, we adopt $R_c = 150$ kpc as characteristic of the Local Group,
although we use a slightly higher value of $n_{-3} = 0.3$ cm$^{-3}$ 
for the central density (Pildis \& McGaugh 1996).
The correction factors in $\S$4.2 lead to high values for 
the upper limit above, viz. $\varphi^{lg} < 2\times 10^6$ ($T_e = 0.34$ keV)
and $\varphi^{lg} < 5\times 10^6$\phoflux\ ($T_e = 0.22$ keV) at
$\Theta_c = 1.17$. 

The Compton $y$ limit does not provide a strong constraint. The quadrupole 
anisotropy from {\sl COBE} (Bennett\etal\ 1994)
is more restrictive by a factor of ten on the product $n_{-3} R_c T_e$ 
(Suto\etal\ 1996, eqn. 9) compared to eqn.~\ref{suto}. (For Hickson
groups, Pildis \& McGaugh restrict this product by another order of
magnitude.)

However, for our adopted parameters, the product $n_{-3} R_c T_e$ is typical 
of spiral groups (Pildis \& McGaugh 1996) where $T_e = (0.22,0.34)$ keV.
Furthermore, the expected ionizing flux out to 500 kpc (eqn.~\ref{bremss})
falls in the range $\varphi^{lg} \sim (1-6,0.5-3)\times 10^4\phoflux$.

\section{Discussion}

If the 3C 273 sight line is representative of the present day 
Ly$\alpha$ forest, the 
Galaxy and M31 ($\varphi^\star$) dominate the cosmic ionizing field 
$\Pcos$ out to at least 700 kpc along the polar axis. This
is an order of magnitude smaller than the scale of typical $L_*$ 
galaxy separations at the present epoch. Although,
after orienting the galaxies correctly with
respect to the supergalactic plane, we find that while the
Galaxy-LMC and M31-M33 pairs may experience significant
levels of mutual ionization, this is not expected for the Galaxy-M31
pair.

The galactic ionizing fields are highly elongated for our
assumed dust opacity $\tau \approx 3$ (eq.~\ref{uvdisk}). 
Only a small fraction of the Local Volume ($5-10$\%) is influenced
by the stellar UV field.
At large galactocentric radius, there exists a toroidal shadow region
close to the galactic plane (BM). Within this region, the galactic 
coronal emission, $\varphi^{gc}$, could well exceed \Pcos.

The COBE sky temperature measurements permit a Local Group corona which 
produces detectable levels of UV flux.  The cooling time of the 
gas is $\sim4\times 10^9$ yrs and therefore would need to be 
replenished (\eg ram pressure stripping, galactic winds).
If confirmed, there are some interesting consequences for $\Pcos$:
(i) attempts to measure the truly {\it cosmic} UV background directly 
(\eg Henry 1996) cannot be made from our vantage point;
(ii) constraints on $\Pcos$ from \Ha\ emission (\eg Vogel\etal\ 1995)
require HI clouds which are not associated with galaxy groups.

More than half of all galaxies reside within small groups, and spirals 
dominate these groups. Could pervasive coronal emission in galaxy
groups explain truncated HI disks 
in spiral galaxies (cf. Maloney 1993)? Could this same
medium explain the unusual ionization conditions observed
along sight lines towards extragalactic sources (Sembach\etal\ 1998)?

Several authors have suggested that some fraction of HVCs, particularly
the more compact clouds, are external to the Galaxy (Blitz\etal\ 1996; 
Spergel\etal\ 1996; Sembach\etal\ 1995, 1998). \Ha\ non-detections towards
these clouds would argue {\it for} the extragalactic model, and 
{\it against} the pervasive intragroup corona, in which case the clouds 
could be used to set hard limits on the cosmic ionizing field.

 { \bf Acknowledgments}
We are indebted to P.R. Maloney for use of unpublished results arising 
from collaborations, and to R.B. Tully, G. Da Costa and B. Dorman 
for their insights.

\appendix 
\section{The Local Group}

The Local Group has at least 40 members. Some of the outliers may not be 
bound to the group. The supergalactic 
coordinates (SGX,SGY,SGZ) are deduced from the NASA/IPAC Extragalactic 
Database.
\begin{table}
\begin{tabular}{llcccc}
name~~~~~~~~~~~~~~~~~~~~~		& type~~~~~~~   &    ~~SGX~~  &    ~~SGY~~  &    ~~SGZ~~  &   ~~$\log L_B$~~ \\
\\
M31             & S      &    0.68 &   -0.30 &    0.17 &   10.48 \\ 
Galaxy          & S      &    0.00 &    0.00 &    0.00 &   10.30 \\ 
M33             & S      &    0.71 &   -0.43 &    0.00 &    9.78 \\ 
LMC             & Irr    &   -0.03 &   -0.02 &   -0.03 &    9.48 \\ 
SMC             & Irr    &   -0.04 &   -0.04 &   -0.01 &    8.85 \\ 
IC 10           & Irr    &    0.58 &   -0.06 &    0.19 &    8.70 \\ 
NGC 3109        & Irr    &   -0.66 &    0.59 &   -0.89 &    8.48 \\ 
NGC 205         & E      &    0.69 &   -0.30 &    0.17 &    8.48 \\ 
M32             & E      &    0.68 &   -0.31 &    0.17 &    8.30 \\ 
NGC 6822        & Irr    &   -0.19 &   -0.21 &    0.44 &    8.30 \\ 
WLM             & Irr    &    0.13 &   -0.93 &    0.13 &    8.30 \\ 
NGC 404         & E      &    2.15 &   -1.15 &    0.27 &    8.30 \\ 
NGC 185         & E      &    0.60 &   -0.18 &    0.16 &    8.30 \\ 
Leo A           & Irr    &    0.66 &    1.82 &   -0.93 &    8.30 \\ 
NGC 147         & E      &    0.57 &   -0.17 &    0.16 &    8.00 \\ 
IC 5152         & Irr    &   -0.77 &    0.50 &   -0.01 &    8.00 \\ 
IC 1613         & Irr    &    0.30 &   -0.35 &   -0.62 &    8.00 \\ 
Pegasus         & Irr    &    0.98 &   -1.36 &    0.76 &    7.95 \\ 
Sextans A       & dIrr   &   -0.30 &    0.88 &   -0.80 &    7.90 \\ 
Sextans B       & dIrr   &   -0.09 &    0.94 &   -0.78 &    7.70 \\ 
DDO 210         & dIrr   &   -0.12 &   -0.37 &    0.47 &    7.48 \\ 
1001-27         & dIrr   &   -0.67 &    0.56 &   -0.87 &    7.00 \\ 
Fornax          & dSph   &   -0.01 &   -0.12 &   -0.07 &    6.85 \\ 
DDO 187         & dSph   &   -0.27 &    1.94 &    0.89 &    6.85 \\ 
DDO 155 (GR8)   & dIrr   &   -0.34 &    1.49 &    0.12 &    6.60 \\ 
Sculptor        & dIrr   &   -0.01 &   -0.08 &   -0.01 &    6.60 \\ 
Andromeda I     & dSph   &    0.52 &   -0.29 &    0.14 &    6.30 \\ 
Andromeda II    & dSph   &    0.52 &   -0.24 &    0.13 &    6.30 \\ 
Andromeda III   & dSph   &    0.62 &   -0.31 &    0.14 &    6.10 \\ 
SAGDIG          & dIrr   &   -0.26 &   -0.23 &    0.51 &    6.30 \\ 
Phoenix         & dIrr   &   -0.11 &   -0.39 &   -0.15 &    6.30 \\ 
Leo I           & dSph   &    0.08 &    0.23 &   -0.12 &    6.00 \\ 
Leo II          & dSph   &    0.00 &    0.19 &   -0.13 &    5.90 \\ 
Tucana          & dSph   &   -0.62 &   -0.68 &   -0.01 &    5.78 \\ 
Andromeda IV    & dSph   &    0.53 &   -0.31 &    0.05 &    5.78 \\ 
LGS 3           & dIrr   &    0.46 &   -0.41 &    0.04 &    5.70 \\ 
Sextans         & dSph   &   -0.02 &    0.06 &   -0.05 &    5.70 \\ 
Draco           & dSph   &    0.04 &    0.04 &    0.05 &    5.60 \\ 
Ursa Minor      & dSph   &    0.04 &    0.04 &    0.03 &    5.48 \\ 
Carina          & dSph   &   -0.04 &   -0.02 &   -0.07 &    5.30 \\ 
\end{tabular}
\end{table}

\end{document}